\documentclass[12pt]{article}
\usepackage{graphics}
\usepackage{epsfig}
\usepackage{amssymb,amsmath}
\usepackage{array}

 \makeatletter
\def\alt{\mathrel{\mathpalette\gl@align<}}
\def\agt{\mathrel{\mathpalette\gl@align>}}
\def\gl@align#1#2{\lower.6ex\vbox{\baselineskip\z@skip\lineskip\z@
\ialign{$\m@th#1\hfil##\hfil$\crcr#2\crcr\sim\crcr}}} \makeatother

\begin{document}

\begin{flushright}
BA-09-21

UMD-PP-09-042 \\
%July, 2009
\end{flushright}

\vspace*{1.0cm}
\begin{center}
\baselineskip 20pt

{\Large\bf Observable $ \mathbf{n - \overline{n}}$ Oscillations with New Physics at LHC}

\vspace{1cm}

{\large M. Adeel Ajaib$^{a}$\footnote{ E-mail: adeel@udel.edu},
I. Gogoladze$^{a}$\footnote{E-mail:
ilia@physics.udel.edu\\ \hspace*{0.5cm} On  leave of absence from:
Andronikashvili Institute of Physics, GAS, 380077 Tbilisi, Georgia.},
Yukihiro Mimura$^{b}$
%\footnote{ E-mail: }
and
Q. Shafi$^{a}$}

\vspace{1.0cm}

{%\baselineskip 25pt

\it $^a$Bartol Research Institute, Department of Physics and Astronomy, \\
University of Delaware, Newark, DE 19716, USA \\ \vspace{2mm}
$^b$Department of Physics, University of Maryland, College Park, MD 20742, USA
}

\vspace{.5cm}

\vspace{1.5cm} {\bf Abstract}
\end{center}

We explore extensions of the MSSM in which TeV scale vector-like multiplets can mediate observable
$\bf n-\overline{n}$ oscillations, without causing conflict with the proton decay experiments,
%A U(1) symmetry, later extended to also implement the Froggatt-Nielsen mechanism,
with a U(1) symmetry playing an important role.
%It forbids $\Delta B=1$ operators while $\Delta B=2$ are permitted.
The colored vector-like particles, in particular, may be found at the LHC through some decay modes
arising from %baryon number violating couplings.
their direct couplings to quarks.

\thispagestyle{empty}
%\bigskip
\newpage

%\addtocounter{page}{-1}

%%%%%%%%%%%%%%%%%%%%%%%%%%
%\baselineskip 36pt
% Main body
%%%%%%%%%%%%%%%%%%%%%%%%%%
\baselineskip 18pt
%%%%%%%%%%%%%%%%%%%%%%%%%%

\section{Introduction}

Low energy supersymmetry (SUSY) provides one of the most compelling extensions of the standard model (SM).
In particular, the minimal supersymmetric standard model (MSSM) with unbroken `matter' parity offers
an attractive resolution of the gauge hierarchy problem, delivers a compelling cold dark matter candidate,
and nicely implements gauge coupling unification. It also predicts that the lightest CP-even Higgs mass,
assuming TeV scale soft SUSY breaking parameters, cannot be much heavier than about 130 GeV.
It is gratifying to note that the MSSM
predictions will be tested at the soon to be launched Large Hadron Collider (LHC).

Possible extensions of the MSSM have been frequently discussed in the literature.
In a recent paper \cite{Babu:2008ge} it was shown that by adding new TeV scale vector-like particles,
it is possible to increase the upper bound on the mass of the lightest CP-even Higgs boson to 160 GeV,
provided there exists a
direct coupling between the new vector-like particles and the MSSM Higgs fields.
To retain gauge coupling unification it is helpful that the new particles comprise  the same particle  content
as the complete SU(5) multiplets such as $\bf 5+\overline{5}$ and $\bf 10+\overline{10}$.

Since LHC is a hadron collider,
it is expected that the colored components of these vector-like fields can be produced there.
In general, these colored particles will couple to  quarks and leptons,
which makes the discussion of the creation and decay modes of the vector-like matter
quite interesting \cite{Mohapatra:2007af}.
In addition to the direct creation of the new particles in LHC,
the new couplings with the new colored particles can generate, as we will see, exotic phenomena related to baryon number violation.
Indeed, it has been noted in the past that  $\bf n-\overline{n}$ oscillations
\cite{Kuzmin:1970nx,Glashow,Mohapatra:1980qe,Riazuddin:1981sn,Lazarides:1986jt,Chacko:1998td,Goity:1994dq,
Mohapatra:2009wp,BaldoCeolin:1994jz}
can be experimentally accessible if there exist TeV scale colored particles with
appropriate baryon number violating couplings to the known particles \cite{Chacko:1998td}.

 Baryon number $B$ is conserved  to all orders in perturbation theory
 in the standard model, up to Planck scale suppressed operators.
On the other hand, a naive SUSY extension of the  SM leads to  rapid nucleon decay via
renormalizable couplings.
This disaster is usually prevented by introducing a discrete $Z_{2}$ `matter' (or $R$) parity.
This version of the MSSM with unbroken matter parity allows  nucleon decay to proceed via Planck
scale suppressed dimension-five operators.
With the current proton lifetime limit $\tau _{p}$ $\gtrsim 5\times 10^{33}$ yrs.,
this implies that the dimensionless coefficients accompanying these operators must be tiny,
of order $10^{-7}$ or less.
The coefficients can usually be arranged to be tiny by a flavor structure
similar  to the small Yukawa couplings for the first generation.
However, the flavor suppression is not enough
if the non-renormalizable operator is generated via the exchange of the vector-like fields
whose masses are much lower than the Planck scale.
Besides, dimension-six nucleon decay operators generated via the TeV scale vector-like fields
also become dangerous, and thus, irrespective of SUSY or non-SUSY models,
the dangerous nucleon decay operators
should be forbidden by some suitable symmetry  \cite{Babu:2003qh,Dreiner:2006xw}.
Nevertheless, it seems important to investigate whether baryon number violation can
manifest itself through other processes, such as
$\bf n-\overline n$ oscillations
mediated by  new particles in the TeV mass range and which may be
experimentally accessible at the LHC.

In this paper, we consider a  framework to discuss  baryon number violation in which
a U(1) symmetry forbids  $\Delta B=1$ operators while $\Delta B=2$ operators are permitted.
In such a  framework, $\bf n-\overline{n}$ oscillation operators are allowed
and they arise via exchange of  new TeV scale vector-like fields,
while  rapid nucleon decays are avoided.
We estimate the size of the oscillation time by imposing a familiar flavor symmetry,
and show that there is an upper bound on the oscillation time for a given vector-like
mass scale.
With the vector-like masses  at 1 TeV,
the bound lies just above the current experimental bound and  can be
tested in future experiments.
The allowed couplings among the vector-like particles and the MSSM fields
can be observed, as we will show, at the LHC.

This paper is organized as follows:
In section 2,
we introduce a U(1) symmetry to forbid
$\Delta B %=\Delta L
=1$ nucleon decay processes,
but allow $\Delta B=2$ operators responsible for $\bf n-\overline n$
oscillations.
In section 3,
we show how the $\bf n-\overline n$ oscillation operator
is generated by means of the TeV scale vector-like fields.
In section 4,
a flavor symmetry is introduced
to estimate the order of the oscillation time,
and we show that $\bf n-\overline n$ oscillations may be observed in
the next generation experiments.
In section 5, we study the implication of the TeV scale
vector-like matter fields in LHC experiments.
Section 6 summarizes our  conclusions.

\section{Symmetry to forbid $\Delta B=1$ nucleon decay operators}

The conservation of the baryon number $B$ and the lepton number $L$
is an important conventional law in high energy physics,
and its violation can play a role to construct a model beyond the standard model.
The lepton number conservation can be violated by the right-handed neutrino Majorana masses,
and it can explain the tiny masses of the neutrinos by the seesaw mechanism \cite{seesawI}.
On the other hand, the baryon number violation has not been observed,
and it is important to search a possible baryon number violating phenomenon.
The particles in the standard model are a collection from the observations below
a few hundred GeV,
and it is surprising that the renormalizable Lagrangian which is written by the
collection of the particles has an accidental baryon number symmetry.
This fact implies that the baryon number symmetry may be related to
an underlying theory beyond the standard model.

In MSSM, a baryon number violating term is allowed by the SM gauge symmetry
even in the renormalizable level,
and thus $Z_2$ parity called $R$-parity is usually imposed
to avoid rapid nucleon decays.
The $R$-parity can be considered a discrete subgroup of the U(1)$_{B-L}$ symmetry \cite{Mohapatra:1986su},
and in this sense,
the symmetry is also related to the lepton number symmetry,
whose violation is responsible for the neutrino Majorana mass.
The way to obtain the discrete symmetry from the baryon and lepton number symmetry
is not unique,
and in this section,
we will investigate the generalization of the discrete symmetry
to a linear combination of the baryon and lepton number symmetry.

We will add vector-like fields to the MSSM particle content,
the baryon symmetry can then be broken in general,
and a symmetry has to be imposed to keep the nucleon adequately stable.
The heavy new fields can be integrated out, and the integration
does not change the symmetry of the Lagrangian.
Therefore, when we consider the general Lagrangian including non-renormalizable couplings
in terms of the MSSM fields,
we can study the symmetry of the system independent of the choice of the vector-like fields.

Let us start from the following superpotential terms which respect $R$-parity:
\begin{equation}
W = q u^c H_u + q d^c H_d + \ell e^c H_d + \ell \nu^c H_u  + \mu H_u H_d +
M_P \left(\frac{S}{M_P}\right)^m \nu^c \nu^c,
\end{equation}
where $q,u^{c},d^{c},\ell ,e^{c},\nu ^{c}$ are the quark and lepton superfields, $H_{u}$ and $H_{d}$
are the Higgs superfields,  $S$ is an MSSM singlet whose vacuum expectation value (VEV) provides the
Majorana mass of the right-handed neutrino $\nu ^{c}$, and $M_{P}$ denotes the Planck scale.
Since there are 9 fields and 6 terms, there are 3 independent U(1) symmetries in the superpotential.
The three symmetries correspond to the hypercharge U(1)$_{Y}$, baryon and lepton number symmetries.
We can also count the $R$-symmetry independently, but it will not play a role in the following discussion.
The MSSM singlet field $S$ carries lepton number $2/m$, which is spontaneously broken when the scalar
component of $S$ acquires a non-zero VEV.

When a non-renormalizable coupling is introduced,
baryon number symmetry is broken in general.
However, there can still remain one linear combination of the baryon and lepton
number symmetry,
even if we add a baryon number violating non-renormalizable term.
Suppose that we allow a non-renormalizable term
$S^n (u^c d^c d^c)^2$.
Then, one can find that a U(1) symmetry remains as shown in Table 1.
\begin{table}
\begin{center}
\begin{tabular}{|c|c|c|c|c|c|c|c|c|c|} \hline
& $q$ & $u^c$ & $d^c$ & $\ell$ & $e^c$ & $\nu^c$ & $h_u$ & $h_d$ & $S$ \\ \hline
$-(nB+mL)/2$ & $-\frac{n}{6}$ & $\frac{n}{6}$ & $\frac{n}{6}$ &
$-\frac{m}2$ & $\frac{m}{2}$ & $\frac{m}{2}$ & 0 & 0 & $-1$ \\ \hline
\end{tabular}
\end{center}
\caption{U(1) charge assignments to forbid rapid nucleon decay.
With $n$ odd and $m$ even, all $\Delta B=\pm1$ operators
are forbidden.}
\end{table}
Here the normalization is chosen so that the charge of $S$ is $-1$.
In general, we can consider the $-(nB+mL)/2$ symmetry even for $n<0$,
and we can discuss what kind of the baryon and lepton number violation
operators are allowed.
If $n\geq0$, the $\Delta B =-2$ operator $S^n (u^c d^c d^c)^2$
is allowed.
When $n\leq0$, the $\Delta B=2$ (non-chiral) operator
$(S^{\dagger})^n (qqd^{c\dagger})^2$ is allowed.
One can find that the exponent of $S$ or $S^\dagger$
is $-(n\Delta B+m\Delta L)/2$ or $+(n\Delta B+m\Delta L)/2$.
Then, we obtain the important consequence:
$\Delta B = \Delta L = \pm 1$ operators
(e.g. $qqq\ell$, $u^c d^c u^c e^c$ and $qq u^{c\dagger}e^{c\dagger}$)
are forbidden when $n+m$ is an odd number.
This is because, for example, the charge of $qqq\ell$ operator is $-(n+m)/2$.
Therefore,  $\Delta B=\Delta L=\pm 1$ nucleon decay operators
are forbidden under the U(1) symmetry when $n+m$ is odd.
This result is valid as long as we do not introduce an additional
SM singlet field with half-odd-integer charge under the U(1) symmetry and which acquires a non-zero VEV. Similarly, we find that the
$\Delta B=-1$,  $\Delta L=0$ operator ($u^c d^c d^c$), which is an $R$-parity
violating term, is not allowed  under the U(1) symmetry when $n$ is an odd number.
Finally, the $\Delta B=0$, $\Delta L=1$ $R$-parity violating
operators ($qd^c\ell$ $\ell\ell e^c$,  $\ell h_u$) are forbidden when $m$ is odd.

We note that the choice of $n=-1$ and $m=1$ corresponds to
the case where there is a $B-L$ symmetry and an SM singlet field
with a charge $B-L=-2$.
This is the familiar case that the $R$-parity is obtained from a discrete subgroup
 of the $B-L$ symmetry \cite{Mohapatra:1986su, {Kibble:1982ae}}.

If $\Delta B=\Delta L=\pm 1$ is forbidden by the $-(nB+mL)/2$ symmetry,
all the $R$-parity violating terms cannot be forbidden
since $n+m$ is even if both $n$ and $m$ are odd. We will discuss the  way to forbid $R$-parity violating terms
by forbidding the $\Delta B=\Delta L=\pm 1$ nucleon decay operators later.

Since  global symmetry can be broken by gravity, the U(1) symmetry should be gauged.
The $-(nB+mL)/2$ symmetry is anomalous, and Green-Schwarz mechanism \cite{GS,DSW} is applied.
The Green-Schwarz mechanism can be applied only for one of the linear combinations of
the anomalous U(1) symmetry. Therefore, the other linear combination ($mB-nL$ symmetry)
is global, and gravity can break it. So, it can be natural that there are $\Delta B=2$ terms in Lagrangian
even if $\Delta B=1$ is forbidden in this framework.

It is interesting that $\Delta B =\pm 1$ nucleon decay operators
are all forbidden when $n$ is odd and $m$ is even, while $\Delta B=\pm 2$ operators are allowed. In this case,
the cutoff scale of the non-renormalizable operators,
which is related to the new particle scale, can be much lower than the Planck scale or the grand unified scale.
If we can make the new particle scale down to TeV scale,
the new particles can be created at LHC. In the next section, we investigate the $\Delta B = \pm 2$ operators
and see whether we can lower the new particle scale or not.

\section{Vector-like matter and $\Delta B=2$ operators} %$\mathbf{n-\overline n}$ oscillation}

The operator
$u^{c}\,d^{c}\,d^{c}\,u^{c}\,d^{c}\,d^{c}$
induces $\Delta B=\pm 2$, $\Delta L=0$ transitions and contributes to
$\mathbf{n-\overline{n}}$ oscillations.
The six-quark operator has dimension 9 and therefore, the coupling strength scales as
$G_{\Delta B=2}\sim \frac{1}{M_{\ast }^{5}}$, where $M_{\ast }$ is the scale of new physics.
We propose that this scale can be identified with new vector-like matter fields with masses of a few hundred GeV.
It is well known that one can extend the matter sector of the MSSM while preserving gauge coupling unification
provided the additional matter superfields fall into complete multiplets of any suitable unified group, such as SU(5).
Such extended scenarios with TeV scale matter multiplets are well motivated. Within string theory, for instance,
one often finds `light' (TeV scale) multiplets in the spectrum \cite{Dienes:1996du}, and even within the framework of
GUTs one can find extra complete multiplets with masses around the TeV scale~\cite{Babu:1996zv}.

An important constraint on the GUT representations and how many such fields can be present at low energy
($\sim $ TeV) comes from the perturbativity condition, which requires that the three MSSM gauge couplings
remain perturbative up to the GUT scale $M_{G}$. One finds that there are several choices available to satisfy
this constraint: {\it(i)} up to 4 pairs of $({\bf5}+\bar{\bf5})$'s, {\it}{(ii)} one pair of
$({\bf10}+\overline{\bf10})$ or {\it(iii)} the combination, $({\bf5}+\bar{{\bf5}}+{\bf10}+\overline{\bf10})$.
Here all representations refer to SU(5) multiplets. In addition, any number of MSSM gauge singlets can be added
without sacrificing unification or perturbativity. Option {\it(iii)}, along with a pair of MSSM gauge singlets,
are contained in  SO(10) spinor representations $({\bf16}+\overline{\bf16})$.

All of the above cases have previously been studied in the literature. For example, it is clear that
new matter will contribute at one loop level to the CP-even Higgs mass if there is direct coupling between
new matter and the MSSM Higgs fields. The authors in \cite{Moroi:1992zk, Babu:2008ge} conclude that
the mass of the lightest CP-even Higgs mass could be pushed up to 160~GeV, consistent with the perturbativity constraint.

There are constraints on the couplings and masses of new matter fields. The most important constraints are
from the $S$ and $T$ parameters which limit the number of additional {\it}{chiral}\/ generations.
Consistent with these constraints, one should add new matter which is predominantly vector-like.
In the limit where the vector-like mass, $M_{V}$, is much heavier than the chiral mass term
(mass term arising from Yukawa coupling to the Higgs doublets), the contribution to the $T$ parameter
from a single chiral fermion is approximately~\cite{Lavoura:1992np}:
\begin{equation}
\delta T\approx\frac{ N (\kappa v)^2}{10 \pi \sin^2\theta_W m^2_W}\left[ \left(
\frac{\kappa v}{M_V}\right)^2  +O\left( \frac{\kappa
v}{M_V}\right)^4\right], \label{tpar}
\end{equation}
where $\kappa$ is the new chiral Yukawa coupling, $v$ is the
VEV of the corresponding Higgs field, and $N$ counts the additional
number of SU(2) doublets. For instance, $N=3$ when ${\bf10} + \overline
{\bf 10}$ is considered at low scale, while  $N=1$ for the ${\bf5}+\bar{\bf 5}$
case. From precision electroweak data $T\leq 0.06(0.14)$ at 95\% CL for
$m_h=117$ GeV ($300$ GeV) ~\cite{PDG}. We
will take $\delta T<0.1$ as a conservative bound and apply it in our analysis.
We then see from Eq.~(\ref{tpar}) that with $M_V$ around 1~TeV, the
Yukawa coupling $\kappa$ can be $O(1)$. We restrict ourselves by introducing low scale vector-like
matter belonging to the ${\bf5}+\bar{{\bf5}}$ and ${\bf10}+\overline{\bf10}$ dimensional representations
of SU(5) as well as an MSSM singlet field.

The six-quark operators which contribute to $\bf n-\overline n$ oscillations in the SM
are $u^{c}d^{c}d^{c}u^{c}d^{c}d^{c}$, $qqd^{c\dagger }qqd^{c\dagger }$ and
$qqd^{c\dagger }u^{c\dagger }d^{c\dagger }d^{c\dagger }$. Among them, only $u^{c}d^{c}d^{c}u^{c}d^{c}d^{c}$
is holomorphic, and in SUSY theories, it will provide the dominant contribution to $\bf n-\overline n$
oscillations through double gluino and neutralino dressed diagrams (see Figure \ref{f1}) when $m_{\rm{SUSY}}\ll M_{V}$.
The contribution from the holomorphic operator can be estimated as
$G_{\bf n-\overline n} \sim (\alpha_s/4\pi)^2 1/(m_{SUSY}^2 M_V^3)$.
On the other hand, the contribution of non-holomorphic operators generated at tree-level are
estimated as $G_{\bf n-\overline n} \sim 1/M_V^5$. When the vector-like mass $M_{V}$ is
lowered to $m_{\rm{SUSY}}$, the tree-level diagram will dominate even for the holomorphic
contribution using the bilinear term $B_{M_{V}}D_{5}\overline{D}_{5}$,
which is a soft SUSY breaking bilinear mixing term for $M_{V}D_{5}\overline{D}_{5}$.
This contribution is estimated to be
\begin{equation}
G_{\bf n-\overline n}
\sim \frac1{M_V} \left(\frac{B_{M_V}}{(M_V^2 + m_0^2)^2-B_{M_V}^2}\right)^2,
\end{equation}
where $m_{0}$ is a universal soft SUSY breaking scalar mass. Therefore, if $m_{\rm{SUSY}}\sim M_{V}$,
both the holomorphic and non-holomorphic operators can provide comparable contributions to
$\bf n-\overline n$ oscillations.

\subsection{{MSSM}$\, \mathbf{+} \, {\bf5} \,  \mathbf{+}\, \overline{\bf 5}$ and $\mathbf{n-\overline n}$ Oscillations}

The representation  ${\bf5}+\overline{\bf5}$ of SU(5)
decomposes under the MSSM gauge symmetry as follows:%
\begin{equation}
{\bf5} + \overline{\bf5}=L_5\left( {\bf1},{\bf2},-\frac{1}{2}\right) +%
\overline{L}_5\left( {\bf1},{\bf2},\frac{1}{2}\right) +\overline{D}_5
\left( \overline{\bf3},{\bf1},%
\frac{1}{3}\right) +{D}_5\left( {\bf3},{\bf1},-\frac{1}{3}\right).
\end{equation}%

The ${\bf5}+\overline{\bf5}$ alone cannot generate the  effective operator responsible for
$\mathbf{n-\overline n}$ oscillations.
However, it can be done with an additional MSSM singlet field ($N$, $\overline N$).
The additional contribution  to the MSSM superpotential relevant for
 $\mathbf{n-\overline n}$ oscillation  is given by
\begin{eqnarray}
W&=& \kappa_1 qq D_5+
\kappa_{2}  u^c d^c \overline{D}_5+ \kappa _{3} D_5 d^c N
+ \kappa _{4} D_5 d^c \overline{N} \nonumber \\ %+ y^{\nu}L \nu^c H_u
&&+ \frac12 M_{N} N N + %(N N + \overline{N}\, \overline{N}) +
M_{V}\left( \overline{D}_5D_5+\overline{L} _{5}L_{5} + N \overline N\right),
\label{nn11}
\end{eqnarray}%
where, for simplicity, we have taken a common vector-like mass for $\bf5+\overline{\bf5}$
at the GUT scale and omitted family indices. We note that the couplings
$D_{5}d^{c}\nu ^{c}$, $D_{5}u^{c}e^{c}$, and $\overline{D}_{5}q\ell $ are forbidden
for $n+m$ odd by the $-(nB+mL)/2$ symmetry, when $u^{c}d^{c}\overline{D}_{5}$, $qqD_{5}$ couplings
are allowed by the symmetry. We assume that additional fields ($5+\overline5+ N+ \overline N$)
are vector-like even when carrying the $-(nB+mL)/2$ charge.  The Majorana mass $M_N$ is
forbidden by $-(nB+mL)/2$  symmetry but it can be generated for instance from the coupling
like $NNS$ once  $S$ acquires a VEV. If we do not introduce a singlet field with a positive charge,
one of the Majorana terms for $N$ and $\overline{N}$ is absent. We choose only $N$ to have a  Majorana mass term.

\begin{figure}[t]
\centering
\includegraphics[angle=0, viewport = -10 -10 460 200,width=8cm]{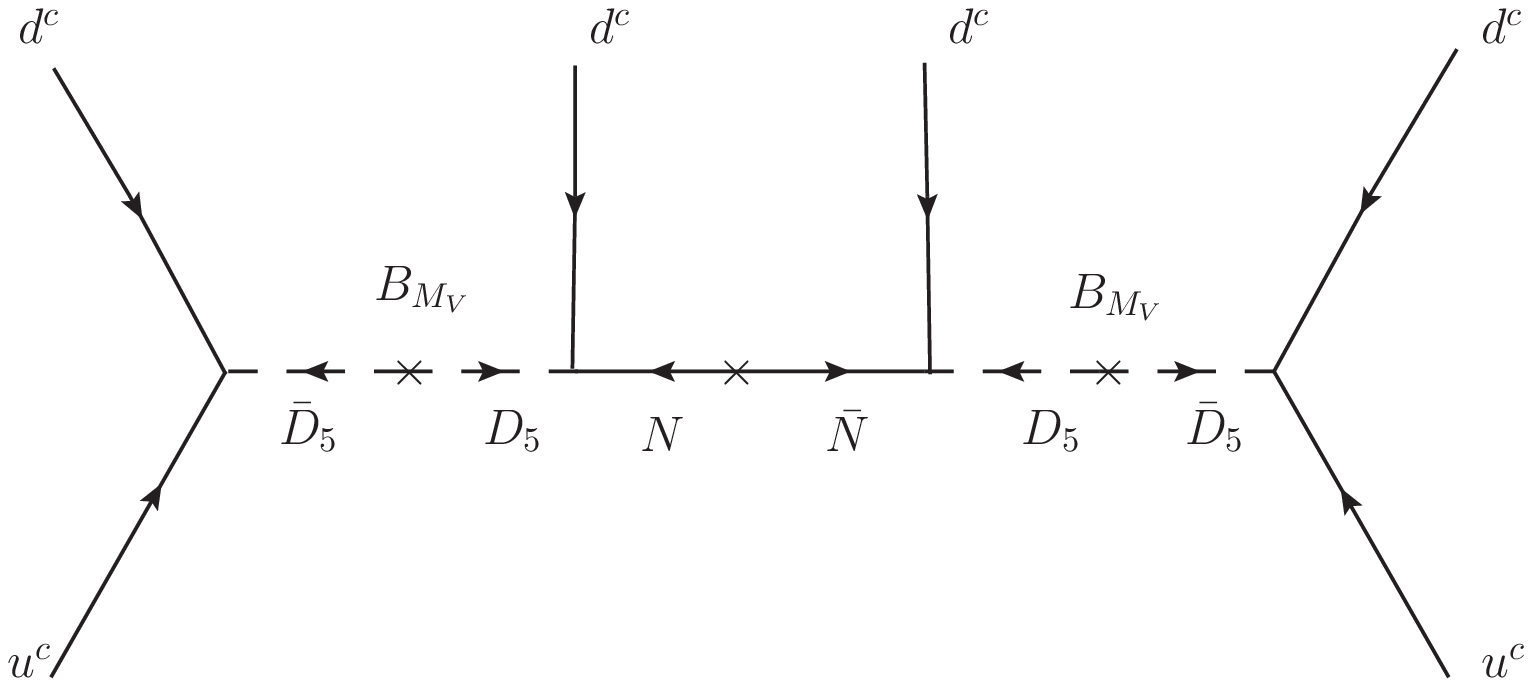}
\includegraphics[angle=0, viewport = -10 -10 460 200,width=8cm]{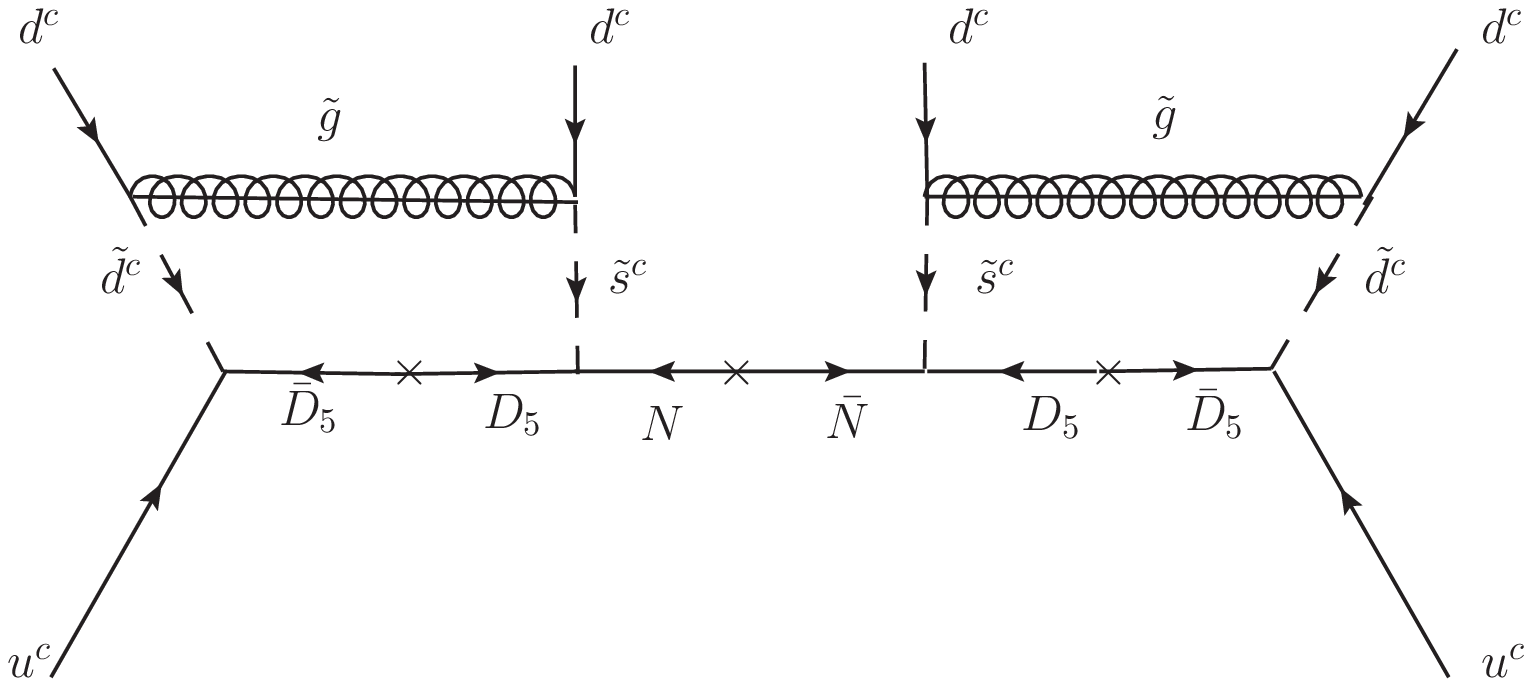}
%\vspace{1cm}
\caption{
Diagrams generating the operator
$(u^c d^c d^c)^2$ via
%$\mathbf{+}$
 $\bf5$ $\mathbf{+}$ $\overline{\bf5}$.
 When the vector-like mass is comparable to the SUSY breaking scale,
 the tree-level diagram (left) dominates instead of the
 gaugino dressed diagram (right).
 } \label{f1}
\end{figure}

\begin{figure}[tbp]
\centering
\includegraphics[angle=0, viewport = -10 -10 460 200,width=9cm]{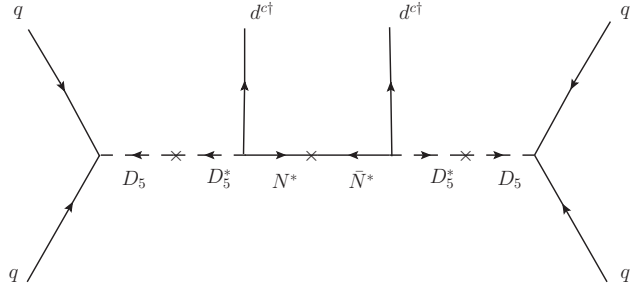}
%\vspace{1cm}
\caption{Diagram which generates the operator $(qqd^{c\dagger})^2$
via exchange of
%$\mathbf{+}$
$\bf5$ $\mathbf{+}$ $\overline{\bf5}$.} \label{f1-2}
\end{figure}

In Figures \ref{f1} and \ref{f1-2}, we show
how the $\bf n-\overline n$ oscillation operators
are generated through the vector-like matter fields.
Since our goal is to relate $\bf n-\bar n$ oscillations to  LHC physics, we assume that
$M_{V}$ is $\sim $ TeV. A small $M_{V}$ compared to the Planck scale may be related to the MSSM $\mu $ problem.
In the literature there exist several mechanisms for explaining why $\mu $ is  $\sim 100$ GeV,
and so we will not address this issue here. To a good approximation we can assume $B_{M_{V}}\approx M_{V}^{2}$,
and for simplicity  set $M_{V}>m_{0}$. With this assumption
  the strength of $\bf n-\overline n$ oscillations can be written as follows,
\begin{eqnarray}
G_{\bf n-\overline n} \simeq
(\kappa_1^2+\kappa_2^2) \left[-\frac12
%(\kappa_3^2+\kappa_4^2)\frac{M_N}{M_V}
\kappa_4^2\frac{M_N}{M_V}
+ \kappa_3\kappa_4
\right] \frac{1}{M_V^5}.
%\frac{(\kappa _{5} \kappa_5^{\prime})^2}{M_R} \frac{1}{(M_V)^4}.
\label{eq1}
\end{eqnarray}
We note that if $M_{N}\gg M_{V}$, one of the mass eigenstates of $N$ and $\overline{N}$ may have a small mass,
and should not be integrated out. This  corresponds to the first term in the bracket.

The current experimental bound on $\bf n-\bar n$ oscillation,
$\tau_{n-\bar n}\geqslant 0.86 \times 10^8$ s \cite{BaldoCeolin:1994jz},  implies an upper limit
\begin{eqnarray}
G_{\bf n-\overline n} \leqslant 3 \times 10^{-28}\ \mbox{GeV}^{-5}.
\label{eq2}
\end{eqnarray}
The bound on the dimensionless coupling can be expressed %assuming $M_V \simeq 10^3$ GeV
%and $M_N \gg M_V$
as follows:
\begin{eqnarray}
%3 \times 10^{-16}\geqslant
%\frac{(\kappa _{5} \kappa_5^{\prime})^2}{M_R}
%
%(\kappa_1^2+\kappa_2^2) \kappa_3 \kappa_4
M_V^5 G_{\bf n-\overline n}
\leqslant \left(\frac{M_V}{1\ {\rm TeV}}\right)^5
\times 3 \times 10^{-13}.
\label{eq3}
\end{eqnarray}
From Eqs. (\ref{eq3}) and (\ref{eq1}), with comparable magnitudes for the couplings $\kappa_i $, we find that
$\kappa _{i}\sim 10^{-3}-10^{-4}$.
In Section 4 we attempt to understand the strengths of these couplings through
the Froggatt--Nielsen mechanism \cite{FrNl}.

\subsection{{MSSM}$\,\mathbf{+}\, {\bf10} \,\mathbf{+}\, \overline{\bf 10}$ and $\mathbf{n-\overline n}$ Oscillations}

The representation  ${\bf10}+\overline{\bf 10}$ of SU(5)
decomposes under the MSSM gauge symmetry as follows:%
\begin{eqnarray}
{\bf10}+\overline{\bf10} &=&Q_{10}\left( {\bf3},{\bf2},\frac{1}{6}\right) +%
\overline{Q}_{10}\left( \overline{\bf3},{\bf2},-\frac{1}{6}\right) +\overline U_{10}\left(
\overline{\bf 3},{\bf1},-\frac{2}{3}\right) +{U}_{10}\left( {\bf3},{\bf1},\frac{2}{3}%
\right)  \nonumber \\
&&+\overline E_{10}\left( {\bf1},{\bf1},1\right) +{E}_{10}\left( {\bf1},{\bf1},-1\right).
\end{eqnarray}%
It was pointed out in ref. \cite{Babu:2008ge} that the light CP-even Higgs mass
in this case can be as large as 160 GeV.
The $\bf 10+\overline{\bf10}$ alone cannot generate the effective operator responsible
for $\bf n-\overline{n}$ oscillations.
However, analogous to the previous subsection, it can be generated
if we introduce an additional MSSM singlet field $(N,\overline N)$,
and the additional contribution to the MSSM superpotential which is relevant for $\bf n-\overline{n}$
oscillations is given by
\begin{equation}
W=\kappa _{10}  d^c s^c \overline U_{10}+
\kappa _{10}^{\prime }{U}_{10}u^c N+
\kappa _{10}^{\prime \prime}{U}_{10}u^c \overline N
%\nu^c  + y^{\nu}_{ij}L_i\nu^c H_u
%+ M_{R_{ij}} \nu^c_i\nu^c_j +
+M_{V}\left( \overline{Q}_{10}Q_{10}+\overline{U} _{10}U_{10} + \overline{E} _{10}E_{10}
+N\overline N\right),
\label{nn12}
\end{equation}%
where, for simplicity, we have assumed a common vector-like mass. %(at the GUT scale $M_G$).

%
%As we will show,
Due to color anti-symmetricity,
the effective operator at tree-level leads to $\mathbf{\Lambda -\overline{\Lambda }}$
rather than $\mathbf{n-\overline{n}}$ oscillation.
However, the $\mathbf{n-\overline{n}}$ oscillation operator can be generated
through radiative corrections via a flavor changing interaction \cite{Goity:1994dq}.
More precisely, we note that the gluino dressing diagram in the previous
subsection also needs a flavor changing interaction in the gluino-squark-quark
vertex due to the total anti-symmetry of the color indices.
When the SUSY breaking bilinear masses are inserted,
we do not need  flavor changing neutral current (FCNC) effects in the diagram for the previous subsection.
However, we do need flavor changing effects induced by radiative corrections even in the case of
SUSY breaking mass insertion
when the $d_{i}^{c}d_{j}^{c}\overline{U}_{10}$ coupling is used,
where $i,j$ are the anti-symmetric generation indices.
Therefore, in this case, the effective $\mathbf{n-\overline{n}}$ transition
operator is rather suppressed compared to the case of vector-like matter fields
in the previous subsection.
We note that non-holomorphic operators are not generated with just
$\bf 10+\overline{10}$ vector-like matter.

\begin{figure}[t]
\centering
\includegraphics[angle=0, viewport = 0 -10 450 200,width=9cm]{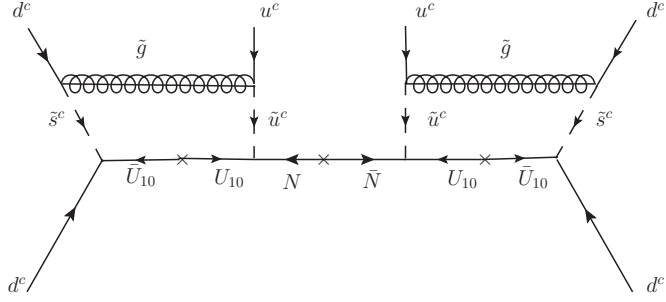}
%\vspace{1cm}
\caption{
Generating $(u^c d^c d^c)^2$
by means of
%$\mathbf{+}$
$\bf10$ $\mathbf{+}$ $\overline{\bf10}$.
Since the $\overline U_{10} d^c_i d^c_j$ coupling is antisymmetric in the flavor
indices,
the $\bf n-\overline n$ transition operator requires radiative corrections
and is suppressed by FCNC. %flavor changing neutral current (FCNC).
} \label{f2}
\end{figure}

\subsection{{MSSM}$\,\mathbf{+}\, {\bf5} \, \mathbf{+}\, \overline{\bf5}\, \mathbf{+}\, {\bf10}\, \mathbf{+}\, \overline{\bf10}$ and $\mathbf{n-\overline n}$ Oscillations}

In this section we will consider extra vector-like matter belonging
to the representations ${\bf5}+ \overline {\bf5} + {\bf10} + \overline{\bf 10}$  of
SU(5).
\begin{figure}[tbp]
\centering
\includegraphics[angle=0, viewport = 0 -10 450 200,width=9cm]{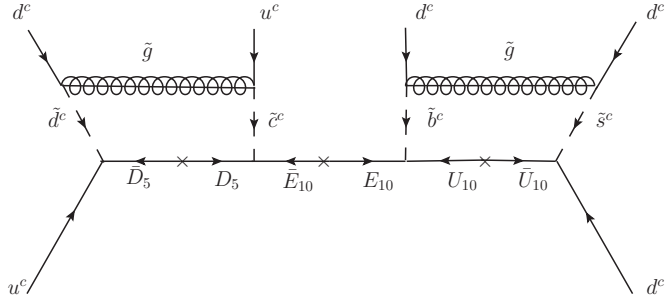}
%\vspace{1cm}
\caption{
Diagram to generate the operator $u^c u^c d^c d^c d^c d^c$
by means of
%$\mathbf{+}$
$\bf 5$ $\mathbf{+}$ $\overline{\bf5}$ $\mathbf{+}$ $\bf10$ $\mathbf{+}$
$\overline{\bf 10}$.
%
%Since the $\overline U_{10} d^c_i d^c_j$ coupling is antisymmetric in the flavor
%indices,
%the $\bf n-\overline n$ transition operator requires radiative corrections
%and are suppressed by FCNC.
%
} \label{f3}
\end{figure}
The MSSM superpotential acquires the following additional contribution
\begin{eqnarray}
W&=&\kappa_1 d^c u^c \overline D_5+ \kappa _{2}{D}_{5}u^c \overline E_{10} +
 \kappa _{3}{E}_{10} d^c {U}_{10}+
 \kappa _4 \overline{U}_{10} d^c (s^c+b^c)  \nonumber
 \\ &+&M_{V}\left( \overline{Q}_{10}Q_{10}+\overline{U}%
_{10}U_{10}+\overline{E}_{10}E_{10} +
\overline{L}_{5}L_{5}+\overline{D_{5}} D_{5}\right).\label{dd666}
\end{eqnarray}%

The holomorphic and non-holomorphic $\bf n-\overline{n}$ oscillation operators are generated as shown in Figure 4
for the holomorphic operators. Again, since we are using the coupling $d_{i}^{c}d_{j}^{c}\overline{U}_{10}$,
we need FCNC %flavor changing neutral currents
to generate $\bf n-\overline{n}$ oscillations as previously discussed.
Thus the operators are suppressed due to the experimental bound on the right-handed quark FCNCs.
%flavor changing neutral currents.

In order to saturate the bound in Eq. (\ref{eq3}), one has to assume that the dimensionless
coefficients in Eq. (\ref{dd666}) are given by
 \begin{eqnarray}
\kappa _{1} \simeq\kappa_2 \simeq\kappa_3 \simeq\kappa_4  \simeq  10^{-3}.
\label{eq1-1}
\end{eqnarray}
These relatively small parameters can be understood within the framework of an anomalous $U(1)_{A}$ flavor symmetry.

\subsection{$\Delta B = \Delta L = 2$ operators}

Before including the flavor structure, we briefly discuss other possible $\Delta B=2$ operators
allowed by the symmetry. In addition to $\Delta B=2$, $\Delta L=0$ operators, $\Delta B=\Delta L=2$ operators
can be allowed under the $-(nB+mL)/2$ symmetry. The $\Delta B=\Delta L=2$ operators (typically $(qqq\ell )^{2}$)
are responsible for $\mathbf{H-\overline{H}}$ (hydrogen-anti hydrogen) oscillations, and double nucleon decays
(e.g. $pp\rightarrow e^{+}e^{+}$) \cite{Mohapatra:1982aj}. The most stringent bound on these operators comes
from the limit on the lifetime for the double proton decay,
$\tau_{pp} \agt 10^{30}$ years \cite{PDG}.
This is interpreted
as $\tau_{\bf H-\overline H} \agt 10^{17}$ years.
This bound implies that $G_{\bf H-\overline H} \alt 3\times 10^{-26}$
GeV$^{-8}$ \cite{Mohapatra:1982aj}.
 If there are vector-like matter fields $\bf 5+\overline{\bf5}+\bf 10+\overline{\bf10}$,
it is possible to generate $\Delta B=\Delta L=2$ operators, similar to the $\mathbf{n-\overline{n}}$ operators.
If the vector-like masses are around 1 TeV, the couplings accompanying these operators should be small,
but not necessarily tiny, compared to the $\mathbf{n-\overline{n}}$ operators. Depending on the choice of
the fermion couplings with the vector-like matter, it is possible for the double proton decay to be observed
instead of $\mathbf{n-\overline{n}}$ oscillations.
With the decay modes to positrons more suppressed than the ones to muons due to flavor suppression,
$pp \to \mu^+\mu^+$
will be a possible observable mode.

\section{ Anomalous $U(1)_A$ Flavor Symmetry and $\bf n-\overline n$ Oscillations}

As we know the Yukawa couplings related to the first and second generations are suppressed and the
inter-generation mixing in the quark sector is small. Anomalous flavor $U(1)_A$ symmetries are often considered
to explain this structure \cite{lfU1}. The new couplings in Eq.(\ref{nn11}) involving quarks, leptons and
vector-like fields can  also be  controlled by this anomalous flavor symmetry.
We will superpose the $U(1)_A$ flavor symmetry on the $-(nB+mL)/2$ symmetry discussed in section 2.

In the string motivated theories, there are towers of vector-like fields at the string/Planck scale.
When those vector-like fields are integrated out, all possible non-renormalizable couplings allowed by
$U(1)_A$ symmetry are generated \cite{FrNl}.
A general form of the superpotential which can explain the fermion
masses and mixing hierarchy through the Froggatt--Nielsen
mechanism  %\cite{FrNl}
has the form
\begin{eqnarray}\label{superP1}
W&=&
y_{ij}^u q_i u_j^c H_u \left(\frac{S}{M_{st}}\right)^{n^u_{ij}}+
y_{ij}^d q_id_j^c H_d \left(\frac{S}{M_{st}}\right)^{n^d_{ij}}\nonumber \\
&+& y_{ij}^e \ell_i e_j^c H_d \left(\frac{S}{M_{st}}\right)^{n^e_{ij}}
 + y_{ij}^\nu \ell_i \nu^c_j H_u \left(\frac{S}{M_{st}}\right)^{n^\nu_{ij}}\nonumber\\
 &+& y_R{}_{ij}  %{M_{R}}_{ij}
 \nu^c_i \nu^c_j M_{st} \left(\frac{S}{M_{st}}\right)^{m+n^{\nu^c}_{ij}}+\mu H_u
 H_d\, ,
\end{eqnarray}
where $M_{st}$ denotes the string scale, $i,j=(1,\,2,\,3)$ are family indices, ${n^u_{ij}}$,
${n^d_{ij}}$, ${n^e_{ij}}$, ${n^{\nu}_{ij}}$ and $n^{\nu^c}_{ij}$
are fixed by the choice of U(1)$_A$ charge assignments.
The numbers $n_{ij}$  are written as
$n^u_{ij} = n^q_i + n^u_j$, $n^d_{ij} = n^q_i + n^d_j$,
$n^e_{ij} = n^\ell_i+n^e_j$, $n^\nu_{ij} = n^\ell_i + n^\nu_j$,
and $n^{\nu^c}_{ij} = n^\nu_i + n^\nu_j$,
where $n^q_i$, $n^u_i$, $n^d_i$, $n^\ell_i$, $n^e_i$, and $n^\nu_i$
are all positive integers.
The coefficients ($y^u_{ij}$, $y^d_{ij}$ etc.) are the coupling constants
which are all taken to be $\sim 1$. As previously discussed in section 2,
there are two independent U(1)$_A$  symmetries in addition to the hypercharge.
We can choose to gauge one of the linear combinations, while the other linear combination
is explicitly broken by additional terms. Proceeding as in section 2,
we choose the U(1) symmetry given in Table 2, and call this the anomalous U(1)$_{A}$  symmetry.
\begin{table}
\begin{center}
\begin{tabular}{|c|c|c|c|c|c|c|c|c|c|} \hline
& $q_i$ & $u^c_i$ & $d^c_i$ & $\ell_i$ & $e^c_i$ & $\nu^c_i$ & $h_u$ & $h_d$ & $S$ \\ \hline
U(1)$_A$ & $-\frac{n}{6}+n^q_i$ & $\frac{n}{6}+n^u_i$ & $\frac{n}{6}+n^d_i$ &
$-\frac{m}2+n^\ell_i$ & $\frac{m}{2}+n^e_i$ & $\frac{m}{2}+n^\nu_i$ & 0 & 0 & $-1$ \\ \hline
\end{tabular}
\end{center}
\caption{
The U(1)$_A$ charge assignment.
The integers (such as $n^q$, $n^u$ and so on)
to generate the mass hierarchy for quarks and leptons
are given in
 Eq.(\ref{flavor-charge}).
}
\end{table}
If $n+m$ is an odd integer, $\Delta B = \Delta L=\pm 1$ operators are all forbidden.
We assign the integers ($n_i^q$, $n_i^u$, etc.) %to the MSSM fields
such that the observed fermion mass and mixing hierarchies are realized.
%obtained with all Yukawa couplings being order one.
As we will show explicitly, the
expansion parameter $\epsilon= \langle S\rangle/M_{st}$ is naturally of
order $0.2$ in the anomalous U(1) models.

We employ the textures  advocated in Ref.
\cite{bgw,Babu:2003zz}, namely
\begin{eqnarray}\label{massM1}
&&M_u\sim \langle
H_u\rangle
\left(
  \begin{array}{ccc}
 \epsilon^{8-2\alpha}&\epsilon^{6-\alpha}&\epsilon^{4-\alpha}\\
\epsilon^{6-\alpha}&\epsilon^4&\epsilon^2\\
\epsilon^{4-\alpha}&\epsilon^2&1
 \end{array}
\right)
\,,\hspace{1.cm}
M_d\sim \langle H_d \rangle\epsilon
\left(
  \begin{array}{ccc}
\epsilon^{5-\alpha}&\epsilon^{4-\alpha}&\epsilon^{4-\alpha}\\
\epsilon^3 &\epsilon^2 & \epsilon^2\\ \epsilon&1&1
 \end{array}
\right) ,\nonumber  \\
\nonumber \\
%\nonumber \\
&&M_e\sim \langle
H_d\rangle \epsilon
\left(
  \begin{array}{ccc}
 \epsilon^{5-\alpha}&\epsilon^{3}&\epsilon\\
\epsilon^{4-\alpha}&\epsilon^2&1\\
\epsilon^{4-\alpha}&\epsilon^2&1
 \end{array}
\right)
\,,\hspace{1.2cm}
M_{\nu_{D}}\sim \langle H_u \rangle\epsilon^{\gamma+1}
\left(
  \begin{array}{ccc}
\epsilon^{2}&\epsilon & \epsilon \\
\epsilon &1 & 1\\
\epsilon & 1 & 1
 \end{array}
\right) ,\nonumber  \\
\nonumber \\
%\nonumber \\
&&M_{\nu^c}\sim
M_{st} \epsilon^{m+2\gamma}
\left(
  \begin{array}{ccc}
 \epsilon^{2}&\epsilon&\epsilon\\
\epsilon &  1 &1\\
\epsilon &  1 &1
 \end{array}
\right)
 \hspace{.5cm} \Rightarrow \hspace{.5cm}
M_{\nu}^{\rm light}\sim \frac{\langle H_u \rangle^2}{\epsilon^{m-2}M_{st}}
\left(
  \begin{array}{ccc}
\epsilon^{2}&\epsilon & \epsilon \\
\epsilon &1 & 1\\
\epsilon & 1 & 1
 \end{array}
\right),
\end{eqnarray}
with
\begin{eqnarray}
n_i^q = (4-\alpha,2,0), \ n_i^u = (4-\alpha,2,0),
\ n_i^d = (2,1,1), \nonumber \\
 \ n_i^\ell = (2,1,1),
\ n_i^e = (4-\alpha,2,0), \ n^\nu_i = (\gamma+1,\gamma,\gamma),
\label{flavor-charge}
\end{eqnarray}
where $\alpha$ is 0 or 1.
Here $M_u$, $M_d$ and $M_e$ are the up--quark, down--quark, and
the charged lepton mass matrices (written in the basis $uM_uu^c$, $d M_d d^c$,
etc.), $M_{\nu_D}$ is the Dirac neutrino mass matrix, and
$M_{\nu^c}$ is the right--handed neutrino Majorana mass matrix.
The light neutrino mass matrix $M^{\rm light}_\nu$ is derived from the
seesaw mechanism. We have not exhibited order one coefficients in
the matrix elements in Eq. (\ref{massM1}). The expansion parameter
is $\epsilon\sim 0.2$,  and in this case  $\tan\beta \sim 10$.
Note that the texture alone does not fix the integer $\gamma$
%exponent $S$
appearing in $M_{\nu_{D}}$ in Eq. (\ref{massM1}),
%depending to the value of $M_R$ we fix S
to get correct light neutrino masses,
while the Majorana mass scale and the neutrino Dirac Yukawa couplings
depend on the integer $\gamma$.
It was shown in Ref.\cite{Babu:2003zz}
that the above  mass matrix texture  nicely fits the observed fermion masses and mixings.

It is interesting that the U(1)$_A$ suppression factor for the holomorphic operators
is calculated independent of the choice of
vector-like matter fields and their U(1)$_A$ charge assignments.
For example,
the $\bf n-\overline n$ oscillation operator
$u^c d^c d^c u^c d^c d^c$
will be given by  the following expression if we use the choice of parameter in
Eq.(\ref{flavor-charge}) [$2 n^u_1 + 4 n^d_1 = 16-2\alpha$]:
\begin{equation}
\frac{1}{M_V^5} \left(\frac{S}{M_{st}}\right)^{n+16-2\alpha} u^c d^c d^c u^c d^c d^c,
\label{n-nbar1}
\end{equation}
where $M_V$ is the effective mass scale associated with the TeV scale vector-like matter
fields.
Additional suppression factors can be supplied through FCNC % flavor changing neutral currents
as previously mentioned.
It is interesting that the current bound on $\bf n-\overline n$ oscillations,
(See  Eqs. (\ref{eq2}, \ref{eq3})),
\begin{equation}
\epsilon^{n+16-2\alpha} \alt \left(\frac{M_V}{1\ {\rm TeV}} \right)^5 \times 3 \times 10^{-13},
\end{equation}
is easily satisfied if $n$ is positive, and with $M_V \sim 1$ TeV.
Note that $0.2^{18} = 2.6 \times 10^{-13}$.

There is another holomorphic operator responsible for  $\bf n-\overline n$ oscillations,
$(qqqH_d)^2$.
Since the down-type quark $s^\prime$ in $q_2 = (c,s^\prime)$ includes the down quark in the mass eigenstate,
the $\bf n-\overline n$ operator can be estimated as [$2n_1^q+4n_2^q=16-2\alpha$]:
\begin{equation}
\frac{1}{M_V^5} \frac{\langle H_d^0 \rangle^2}{M_V^2}
\sin^4\theta_C\left(\frac{S}{M_{st}}\right)^{-n+16-2\alpha} u\,d\,d\,u\,d\,d,
\label{n-nbar2}
\end{equation}
where $\theta_C$ is a Cabibbo angle $\simeq \epsilon$.
It is interesting to note  that the coefficients of $n$ in the exponent
have opposite signatures between Eq.(\ref{n-nbar1}) and Eq.(\ref{n-nbar2}).
This sign change  comes from the fact that these are $\Delta B = -2$ and $+2$ operators.
As a result, the $\bf n-\overline n$ oscillation time has an upper bound
for a given $M_V$ value which  can be estimated as
\begin{equation}
\tau_{n-\overline n} \alt 10^8\  {\rm sec}\ (\alpha=1), \quad 3\times 10^9\ {\rm sec}\ (\alpha=0),
\end{equation}
when $M_V = 1$ TeV and $\tan \beta =10$.
This bound can be achieved by planning experiments with cold neutron horizontal beam \cite{Mohapatra:2009wp}.

\section{Implications for LHC}

As we have previously mentioned, vector-like matter fields at the TeV scale can be found
at the LHC as they decay into  MSSM matter fields as well as gauginos.
The coupling terms with quarks and leptons in the superpotential are
\begin{eqnarray}
W &\!\!=&\!\! \left(\frac{S}{M_{st}}\right)^{n^d_i + n^d_j -\frac{X_U}2}
\overline U_{10} d^c_i d^c_j
+ \left(\frac{S}{M_{st}}\right)^{n^u_i + n^\nu_j +\frac{X_U}{2} + \frac{n+m}{2}}
 U_{10} u^c_i \nu^c_j \\
 &\!\!+&\!\!
 \left(\frac{S}{M_{st}}\right)^{n^q_i + n^q_j +\frac{X_D}2}
 D_{5} q_i q_j
 +
 \left(\frac{S}{M_{st}}\right)^{n^u_i + n^e_j +\frac{X_D}2+\frac{n+m}{2}}
 D_{5} u^c_i e^c_j
 +
 \left(\frac{S}{M_{st}}\right)^{n^d_i + n^\nu_j +\frac{X_D}2+\frac{n+m}{2}}
 D_{5} d^c_i \nu^c_j
 \nonumber \\
 &\!\!+&\!\!
 \left(\frac{S}{M_{st}}\right)^{n^q_i + n^\ell_j -\frac{X_D}2 -\frac{n+m}{2}}
 \overline D_{5} q_i \ell_j
 +
 \left(\frac{S}{M_{st}}\right)^{n^u_i + n^d_j -\frac{X_D}2}
 \overline D_{5} u^c_i d^c_j. \nonumber
\end{eqnarray}
The U(1)$_{A}$ charges of $\overline{U}_{10}$ and $\overline{D}_{5}$ are taken to be
$-n/3-X_{U}/2$ and $-n/3-X_{D}/2$ respectively, where, $X_{U}$ and $X_{D}$ are integers.
The U(1)$_{A}$ charge assignments of vector-like matter are displayed in Table 3.
All unwanted couplings can be excluded by choosing the integers $X_{Q}$, $X_{U}$, etc. to be even or odd.
\begin{table}
\begin{center}
\begin{tabular}{|c|c|c|c|c|c|c|} \hline
& $Q_{10}$ & $\overline U_{10}$ & $\overline D_{5}$ & $L_5$ & $\overline E_{10}$ &
$\overline N$ \\ \hline
U(1)$_A$ & $\frac{n}3+ \frac{X_Q}{2}$ & $-\frac{n}3- \frac{X_U}{2}$ &
$-\frac{n}3- \frac{X_D}{2}$ & $\frac{n}2+ \frac{X_L}{2}$ &
$-\frac{n}2- \frac{X_E}{2}$ & $-\frac{n}2- \frac{X_N}{2}$ \\ \hline
\end{tabular}
\end{center}
\caption{U(1)$_A$ charge assignments
for the vector-like fields.
The parameters $X_Q$, $X_U$ etc. are all integers.
We assume that the charges of the vector-like pair have the same absolute magnitude
but opposite signature (e.g. the charge of $U_{10}$ is $n/3 + X_U/2$).
}
\end{table}
 We assume that the fields are vector-like also under the U(1)$_{A}$ charge, so that their masses
do not depend on the VEV of $S$. If the exponent of the $S$ field is not an integer,
the corresponding coupling is prohibited. One can check that one of the couplings
$D_{5}qq$ or $\overline{D}_{5}q\ell $ is prohibited if $n+m$ is an odd number.
In this case, therefore,
the dangerous operator $qqq\ell $ is not generated by the exchange of $D_5$ and $\overline{D}_5$.

 Suppose we take $X_{D}=0$, so that the $D_{5}q_{3}q_{3}$ coupling coefficient is $O(1)$,
and the $\overline{D}_{5}t^{c}b^{c}$ coupling is $O(\epsilon )$. In this case, $D_{5}$ can
be considered as a `diquark' whose baryon number is $-2/3$. If $D_{5}$ (scalar) is created at the LHC,
its main decay mode will be $D_{5}\rightarrow \bar{t}\, \bar{b}$. If the diquarks are pair created
in the collider, the associated decay mode is $t\,b\,\bar{t}\,\bar{b}$, and therefore it may happen that the
energy distribution of $t\, b\, \bar{t}\,\bar{b}$ deviates from the standard model prediction.
Even if $X_{D}$ is not equal to zero, either the $D_{5}q_{i}q_{j}$ or $\overline{D}_{5}u_{i}^{c}d_{j}^{c}$
coupling will be large, and $D_{5}$ or $\overline{D}_{5}$ can decay into two quarks.
The analyses of diquark production and decay in LHC can be found in \cite{Mohapatra:2007af}.
In these analyses, the diquark is a color sextet instead of a triplet, and decays into $\bar{t}\,\bar{t}$.

 If we take $X_{D}+n+m=0$, then both  $D_{5}u^{c}e^{c}$ and $\overline{D}_{5}q\ell $ couplings are possible.
The field $D_{5}$ for this case can be considered a `leptoquark', and the main decay mode will be
$D_{5}\rightarrow t\,\tau $. Therefore, the decay process from  pair created leptoquarks is
$D_{5}D_{5}^{\ast }\rightarrow t\,\tau\, \bar{t}\,\bar{\tau}$. This decay can be quite significant since
$\tau\bar{\tau}$ production is suppressed in  hadron colliders. However, in this case,
the $\bf{n-\overline{n}}$ oscillation operators will not be generated by means of  vector-like fields.

\begin{figure}[tbp]
\centering
\includegraphics[angle=0, viewport = 0 -10 450 200,width=9cm]{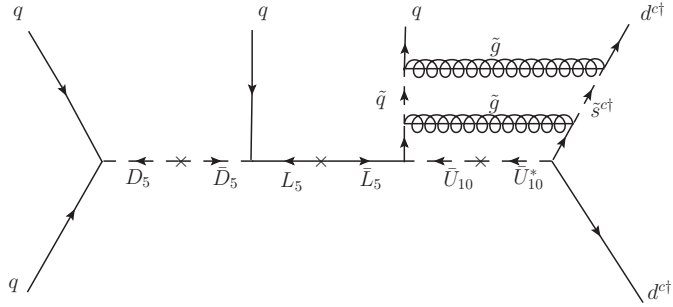}
%\vspace{1cm}
\caption{
Diagram which generates the operator $qqqq d^{c\dagger} d^{c\dagger}$ operator
through exchange of
%$\mathbf{+}$
$\bf 5$ $\mathbf{+}$ $\overline{\bf5}$ $\mathbf{+}$ $\bf10$ $\mathbf{+}$
$\overline{\bf 10}$.
We need radiative corrections to realize the $\bf n-\overline n$ oscillations.
%
%Since the $\overline U_{10} d^c_i d^c_j$ coupling is antisymmetric under the flavor
%index,
%the $\bf n-\overline n$ operator always needs a radiative correction
%and are suppressed by FCNC.
}
\label{f4}
\end{figure}

As it can be seen from Figures \ref{f1}-\ref{f4},  vector-like masses for  uncolored vector-like matters
($L_{5},E_{10},N$) violate  baryon number conservation. If these uncolored matter particles are light and
the colored vector-like particles decay through  uncolored matter exchange, this violation of baryon number
may be observed at the LHC. A possible baryon number violating diquark decay is
$D_{5}\rightarrow b\,b\,D_{5}^{\ast }\rightarrow b\,b\,t\,b$. This possibility is especially interesting if
baryon number violation can be related to the cosmological baryon asymmetry.

Let us consider the constraints on $X_{U}$ and $X_{D}$. The $\overline{U}_{10}$,
$D_{5}$ and $\overline{D}_{5}$ couplings with fermions can generate box diagrams for meson-antimeson mixings
(such as $K$-$\bar{K}$, $D$-$\bar{D}$, and $B$-$\bar{B}$). Note that $\overline{U}_{10}d_{i}^{c}d_{j}^{c}$
couplings are antisymmetric with respect to the flavor index, and thus the meson-antimeson diagram is not
generated at tree-level. If $X_{D}=2$, the $\overline{D}_{10}t^{c}s^{c}$ coupling is $O(1)$, and
the $K$-$\bar{K}$ box contribution through this coupling can be comparable to or even exceed the SM
contribution if the vector-like masses are 1 TeV. Therefore, a large value of $|X_{D}|$ is not favored
if the vector-like matter is to be `light'. The same statement is also true for $X_{U}$. However,
if $X_{U}=X_{D}=0$, then there are flavor suppression factors and the contributions can be much smaller
than the SM contribution. The coupling of the vector-like colored field to the third generation left-handed
quarks is order unity in this case.

As we previously mentioned, the U(1)$_{A}$ suppression factor of the effective holomorphic operators
$u^{c}d^{c}d^{c}u^{c}d^{c}d^{c}$ does not depend on the charge assignments of the
vector-like fields and diagrams which generate this operator.
On the other hand, for the non-holomorphic operators (e.g. $qqqqd^{c\dagger }d^{c\dagger }$)
the U(1)$_{A}$ suppression factor depends on the charge assignments of the vector-like fields.
For example, the U(1)$_{A}$ suppression factor of the operator generated by the diagram in Figure \ref{f4}
is of order $\epsilon ^{16-2\alpha -n-X_{U}}$.
As we have previously mentioned, in addition to the U(1)$_A$ suppression, there is also a FCNC suppression
factor due to the antisymmetricity of the coupling $U_{10}d_{i}^{c}d_{j}^{c}$.
Therefore, if $n$ is a large positive number (which suppresses the holomorphic operator
since the corresponding suppression is $\sim $ $\epsilon ^{16-2\alpha +n}$),
the non-holomorphic operator can become large and in excess of the experimental bound when $X_{U}=0$.
It can be checked that the experimental bound is satisfied due to the flavor suppression factors with
$X_{U}=X_{D}=0$, even if we consider different diagrams to generate the non-holomorphic operators for
$\mathbf{n-\overline{n}}$ oscillations.
It is interesting to emphasize that the coefficient of $n$
in the exponent of the flavor suppression factor can be both positive and negative.
As a result, $\mathbf{n-\overline{n}}$ oscillations can be in the accessible range with
the vector-like masses at TeV scale (accessible at the  LHC),
as we have found in the previous section.
%and there is a source of
%FCNC for the right-handed down-type squarks.

Before closing this section, let us consider the dimension-four $R$-parity violating terms.
The term $u^{c}d^{c}d^{c}$ is forbidden if $n$ is odd,
while $q\ell d^{c}$, $\ell \ell e^{c}$ and $\ell h_{u}$ are forbidden if $m$ is odd.
Thus if $n+m $ is odd to forbid $\Delta B=\Delta L=\pm 1$ operators, we cannot forbid all
dimension-four $R$-parity violating terms. Therefore, $R$-parity should be imposed in our framework.
The $R$-parity of the vector-like fields are chosen to be positive for colored vector-like fields,
and to be negative for uncolored vector-like fields in order to allow the desired couplings for
$\mathbf{n-\overline{n}}$ oscillations.

Due to U(1)$_{A}$ symmetry, the MSSM fermions and the vector-like fields do not mix
(Dirac terms such as $q\overline{Q}_{10}$, $u^{c}U_{10}$ are forbidden) if $n$ is odd, $m$ is even,
and $X_{F}(F=Q,U,D,L,E,N)$ given  in Table 3 are all even. Indeed the couplings
$\overline{U}_{10}d_{i}^{c}d_{j}^{c}$ and $\overline{D}_{5}u_{i}^{c}d_{j}^{c}$
resemble the MSSM $R$-parity violating couplings. It is interesting to note that the decay modes
of the $\overline{D}_{5}$ and $\overline{U}_{10}$ may be confused with the $R$-parity violating
decay of the right-handed squarks at the LHC.
However, since the mixing between $u^{c}$ and $\overline{U}_{10}$ (as well as the mixing between
$d^{c}$ and $\overline{D}_{5}$) is forbidden, this does not contradict with the existence of stable nuclei.

We also need to check that a large neutrino mass term ($\ell \ell H_{u}H_{u}$) is not
generated from the presence of vector-like fields since the $\Delta L=2$ term is allowed by the symmetry
in the framework. If  terms like $\ell q\overline{D}_{5}$, $\ell Q_{10}d^{c}$, $\ell Q_{10}\overline{D}_{5}$,
$\ell L_{5}e^{c}$ etc. are allowed,  unacceptably large neutrino masses can be generated by loop diagrams.
Such terms are all forbidden when we impose $R$-parity as above and
 choose $n,m$ and $X_F$ so as not to
mix the vector-like fields with the MSSM fields. %in the choice above.

As a result, we can conclude that vector-like particles can lie in the TeV range and they can be found
at the LHC without contradicting any of the current experimental bounds. These particles can give rise
to observable $\mathbf{n-\overline{n}}$ oscillations.

We note that the vector-like fields can couple with the MSSM Higgs fields
when we choose $X_Q=X_U$ for example.
Then, the bound of the lightest CP-even Higgs mass can be increased.

Under the anomalous $-(nB+mL)/2$ gauge symmetry,  dimension-six proton decay operators such as
$qqu^{c\dagger }e^{c\dagger }$ are forbidden. Therefore, the string scale, in principle,
can be much lower than the (4 dimensional) Planck scale or even the GUT scale.
The bound on the string scale will come from the general non-holomorphic $\Delta B=2$ operators
(e.g. $(S^{\dagger })^{n+2}qqd^{c\dagger }u^{c\dagger }d^{c\dagger }d^{c\dagger }$).
For the general non-holomorphic operators, the flavor suppression may not work so well.
If the string scale is lowered, the U(1)$_{A}$ gauge boson mass is also lowered.
Then, FCNCs induced by the U(1)$_{A}$ gauge boson \cite{Maehara:1979kf} may have an impact
on flavor changing processes such as $K$-$\bar{K}$, $D$-$\bar{D}$ and $\mu \rightarrow 3e$.
 For these two reasons, the string scale needs to be more than $10^{5}$-$10^{6}$ GeV.

Finally, we note a generalization of the $-(nB+mL)/2$ symmetry.
We have considered the $nB+mL$ symmetry by introducing one SM singlet with charge $nB+mL=2$.
In general, we can take the SM singlet charge to be $nB+mL=K$.
Suppose that $K=4$, $n=4k+2$, and $m=4k^{\prime }+1$ ($k,k^{\prime }$ are integers).
Then, $\Delta B=\Delta L=1$ as well as all the unwanted dimension-four $R$-parity violating
terms are forbidden, and we do not need to introduce a separate $R$-parity.
Since $\Delta B=2$ is allowed, $\mathbf{n-\overline{n}}$ oscillations are still possible.
However, since $\Delta L=2$ is not allowed with this choice, the neutrino only has a Dirac mass.
Surely, proper neutrino masses can be obtained
since the size of the Dirac neutrino mass can be controlled by a choice of the U(1)$_A$ charge.
If $n=4k+1$, and $m=4k^{\prime }+2$, then $\Delta L=2$ is allowed, and the neutrino has a Majorana mass.
However, $\mathbf{n-\overline{n}}$ oscillations are forbidden in this case.
In either case, the vector-like colored fields can be at TeV scale without contradicting nuclei stability,
and they can be created at the LHC.

\section{Conclusion}

We have explored the possible existence of TeV scale vector-like particles, detectable at the LHC,
and which can mediate observable $\mathbf{n-\overline{n}}$ oscillations without creating any conflict
with proton lifetime limits.
We find that a U(1)$_A$ symmetry generated by $-(nB+mL)/2$, a linear combination of baryon
and lepton number, is particularly effective in achieving this scenario. If $ n + m$ is an odd number,
all $\Delta B=\Delta L=\pm 1$ operators are forbidden, as long as there is no SM singlet field with
half-odd-integer charge under the $-(nB+mL)/2$ symmetry, or, if present, such half-odd-integer charged
SM singlet fields do not acquire VEVs. The $\Delta B=1$, $\Delta L=0$ operator is also forbidden if $n$ is odd.
The nucleon is `sufficiently' stable %($\tau _{Nuc}\sim 3\times 10^{32}$ yrs)
and its lifetime satisfies
the current experimental bounds even if the colored vector-like fields have TeV masses.
We have seen that $\mathbf{n-\overline{n}}$ oscillations could be observed in the near future within
the framework of a familiar flavor model for generating quark and lepton mass hierarchies with
vector-like masses at the  TeV scale. The couplings among the quarks and the vector-like fields which
lead to observable $\mathbf{n-\overline{n}}$ oscillations can be tested through the decay modes of these
new fields. This would provide an exciting new window to probe fundamental interactions at the LHC which
violate baryon number conservation.

\section*{Acknowledgments}

We thank Borut Bajc and Rizwan Khalid for useful comments and discussion.  This work is supported in part by the DOE Grant
\#DE-FG02-91ER40626 (I.G. and Q.S.), GNSF grant 07\_462\_4-270 (I.G.),
and NSF grant No. PHY-0652363 (Y.M.).

\end{document}